\documentclass[runningheads]{llncs}
\usepackage[T1]{fontenc}
\usepackage{graphicx}
\begin{document}
\title{Designing Wine Tasting Experiences for All: The Role of Human Diversity and Personal Food Memory}
\titlerunning{Designing Wine Tasting Experiences for All}
\author{Xinyang Shan\inst{1}\and
Yuanyuan Xu*\inst{1}\and
Yuqing Wang\inst{1}\and Tian Xia\inst{2} \and  Yin-Shan Lin\inst{1,3}}
\authorrunning{Xinyang Shan et al.}
\institute{Tongji University, Shanghai 200292, China \and
Shanghai Jiao Tong University, Shanghai 200240, China\\
\and
Northeastern University,Boston, MA 02115\\
\email{xuyuanyuan@alumni.tongji.edu.cn}}
\maketitle          % typeset the header of the contribution
\begin{abstract}
Wine tourism is a rapidly growing sector, valued at \$29.6 billion globally in 2023, with an anticipated annual growth rate of 5.9\% . However, traditional wine tasting practices are rooted in Western cultural norms, limiting their appeal to a diverse global audience. This study explores how human diversity and personal food memory can create more inclusive and engaging wine tasting experiences, particularly for Chinese tourists.  Over 11 months, we conducted field studies in Xinjiang, Shandong, and Tuscany with 23 Chinese participants. Data were collected through observations, audio/video recordings, and follow-up interviews. The Abilities, Necessities, and Aspirations (ANA) framework was applied to identify cross-cultural adaptation challenges. Our findings reveal that 76\% of participants found Western wine terminology difficult to understand, while 83\% preferred describing wine flavors using familiar food-related terms like “pickled radish” and “sun-dried plums”. Additionally, 68\% reported increased confidence and enjoyment when connecting wine flavors to personal food memories. We propose a three-part framework for designing inclusive wine tasting experiences: adapting tasting guides to reflect cultural differences, balancing sensory exploration with social elements, and integrating personal food memory into flavor descriptions .  

This study provides practical guidelines for developing culturally sensitive wine tasting experiences, improving engagement and satisfaction among diverse audiences .  

\end{abstract}

\section{Introduction}

Wine tourism has become a significant part of the global travel industry, with wine tasting at the heart of this growing sector. The global wine tourism market was valued at \$29.6 billion in 2023 and is projected to grow at an annual rate of 5.9\% over the next five years \cite{santos2022towards}. However, traditional wine tasting practices remain deeply rooted in Western cultural norms, which may not resonate with consumers from diverse cultural backgrounds \cite{kastenholz2023wine}.  

Current wine tasting practices are typically designed for Western consumers, emphasizing sensory analysis and technical language to describe flavors \cite{spence2020multisensory}. This creates barriers for consumers from different cultural backgrounds, especially those from China, where drinking culture is more symbolic and socially driven \cite{zhang2020rituals}. Recent works highlight how culturally adapted experience management tools \cite{xu2024design}, AI-supported cross-cultural frameworks \cite{xu2025ai}, emotion-aware interaction design \cite{xu2024utilizing}, and language model applications \cite{shan2024cross} can support more inclusive wine tourism experiences. The role of consumer-perceived value in tailoring such experiences has also been emphasized in business-modeling contexts \cite{xu2024exploring}.

\section{Background and Literature Review}
In Chinese culture, wine is often perceived as a luxury product, valued more for its symbolic meaning than for its sensory qualities \cite{sun2021chinese}. Unlike Western dining traditions, where wine pairing is an important aspect of the meal, Chinese culinary customs involve serving multiple dishes simultaneously without strict pairing guidelines \cite{yang2020flavor}. Cultural differences such as these impact how wine experiences are interpreted and appreciated across global markets \cite{xu2024design}. Additionally, emotion recognition and user-centric technological support can play a role in addressing such cultural gaps \cite{xu2024utilizing}.

Western wine tasting tools, such as the wine flavor wheel and Le Nez du Vin, often fail to resonate with Chinese consumers due to their reliance on Western flavor profiles and sensory terminology \cite{chu2020regional}. Terms like "oak," "minerality," and "tannins" are difficult to translate into Chinese sensory contexts, leading to confusion and disengagement. Typeface and visual language also play a role in shaping user experience and sensory communication across cultures \cite{tian2022study}.

\section{Methodology}
To explore how human diversity and personal food memory influence wine tasting, we conducted a series of field studies over 11 months in three major wine regions: Xinjiang, Shandong, and Tuscany.  

Three distinct wine tasting events were organized. The first event, held at a family-owned winery in Xinjiang, combined a vineyard tour, grape picking, and a traditional lunch paired with five wines. The second event, conducted in Shandong, involved a structured tasting of five wines preceded by a 30-minute introduction to wine terminology. The third event, in Tuscany, included a vertical tasting of six vintages from 2007 to 2017.  

Data were collected using direct observations, video and audio recordings, and post-event interviews lasting 20 to 30 minutes. The analysis was guided by the Abilities, Necessities, and Aspirations (ANA) framework \cite{wu2020examining}.  
\begin{figure}[h]
    \centering
    \includegraphics[width=0.8\linewidth]{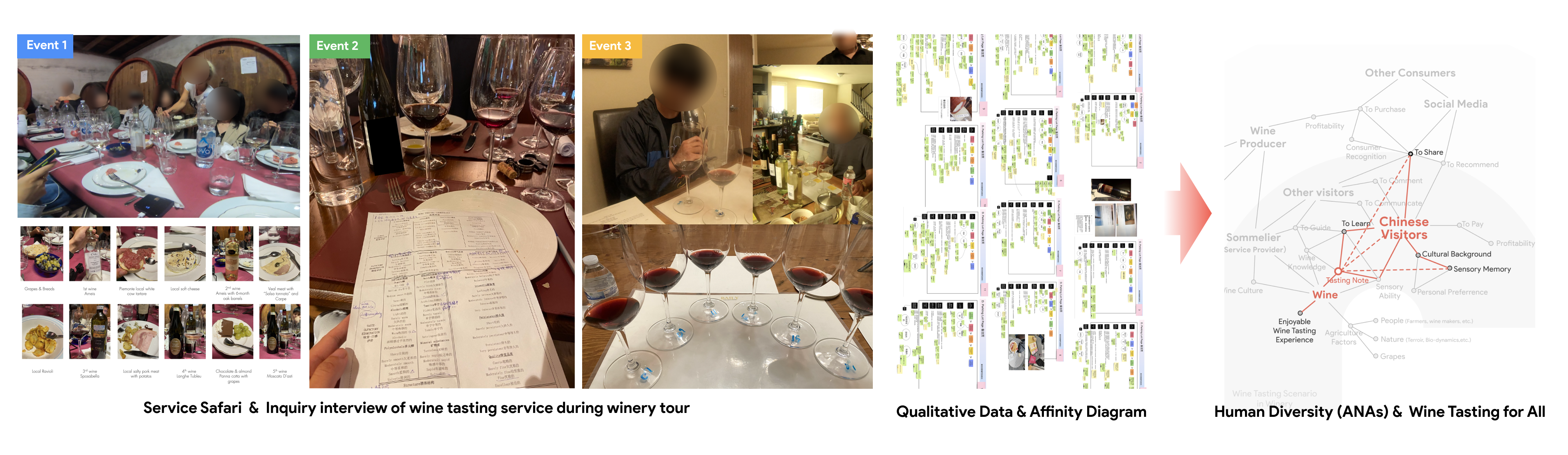}
    \caption{Overview of the study design and methodology.}
    \label{fig:study-design}
\end{figure}

Our findings reveal key patterns in how Chinese participants engage with wine tasting experiences.  

76\% of participants reported difficulty understanding Western wine terminology, such as "oak," "minerality," and "tannins" \cite{chu2020regional}. These terms lacked cultural relevance, making it difficult for participants to interpret sensory characteristics. Even participants with prior wine knowledge showed limited comprehension.  

83\% of participants preferred to describe wine using culturally familiar food-related terms \cite{zhang2020rituals}. For example, participants described red wine aromas as similar to "pickled radish" or "dried hawthorn" and white wine acidity as "sour plum soup." This suggests that adapting sensory language to local culinary references improves engagement.  

68\% of participants reported higher confidence and enjoyment when they were encouraged to describe wine flavors using personal food memory \cite{yang2020flavor}. Connecting wine flavors to familiar foods enhanced sensory engagement and facilitated more active social interaction.  

Participants engaged more actively in informal tasting events (e.g., vineyard-based tastings) than in structured events held in formal venues. Authenticity and comfort increased when the tasting environment resembled social and family-style settings.  

Gender and age differences were observed: female participants showed greater interest in social and symbolic aspects, while younger participants were more open to exploring new flavors. Older participants preferred traditional flavors and described wine using references to Chinese herbal medicine and preserved fruits.

\section{Discussion and Implications}
The results highlight the need to adapt wine tasting experiences to Chinese consumers’ cultural and sensory backgrounds.  

Western wine terminology should be adapted to local culinary references. Terms like "oak" and "tannins" have no direct equivalent in Chinese cuisine. Providing bilingual tasting guides with culturally familiar terms (e.g., "hawthorn," "plum," "green tea") can improve comprehension and comfort \cite{chu2020regional}. Developing a tasting wheel based on Chinese flavor profiles would further support sensory engagement.  

Social interaction is central to Chinese drinking culture. Tasting events should incorporate group-based formats and communal sharing to reflect this social dynamic \cite{zhang2020rituals}. Including storytelling about the vineyard's history and production process can enhance authenticity and emotional connection.  

Personal food memory is a powerful sensory tool. Encouraging participants to describe wine using familiar foods (e.g., "sour plum soup" or "jasmine tea") improves sensory engagement. Training winery staff to integrate culturally relevant terms into tasting descriptions will further enhance this effect \cite{yang2020flavor}.  

Informal and relaxed tasting settings increase participant comfort and perceived authenticity. Balancing structured analysis with casual, social interaction creates a more engaging experience. For instance, combining sensory analysis with informal food pairings could enhance overall satisfaction.  

Gender and age differences should guide the design of tasting experiences. Women showed greater interest in the symbolic and social aspects of wine, while younger participants were more curious about new flavors. Tailoring event formats to these differences can increase overall engagement.  
\begin{figure}[h]
    \centering
    \includegraphics[width=0.8\linewidth]{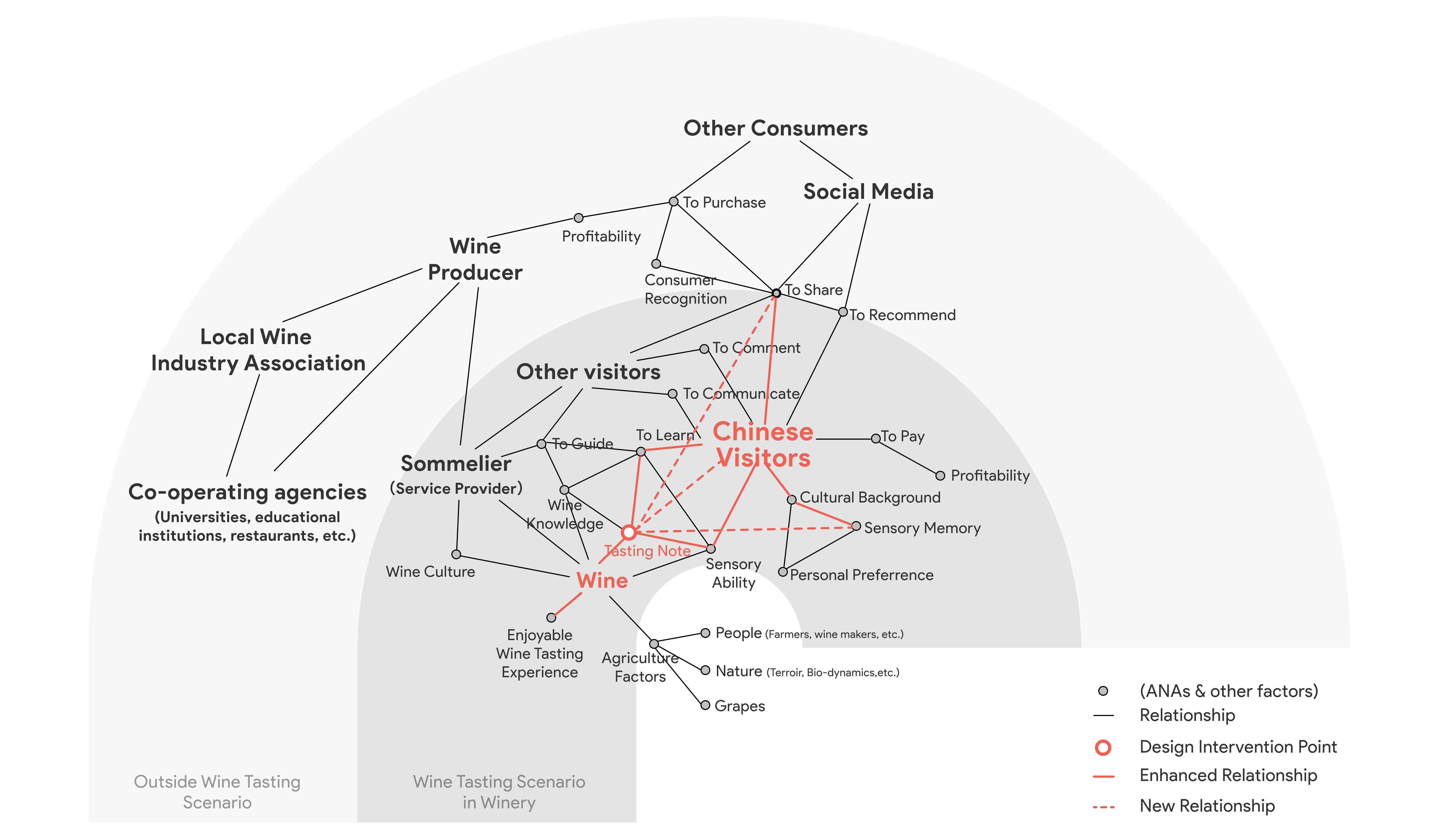}
    \caption{Proposed framework for designing culturally adaptive wine tasting experiences.}
    \label{fig:framework}
\end{figure}

\section{Conclusion}
This study shows that Western wine terminology creates barriers for Chinese consumers. Adapting tasting language to culturally familiar terms improves comprehension and engagement.  

Personal food memory enhances sensory engagement and social interaction. Encouraging participants to describe wine using familiar foods strengthens emotional connection and improves retention.  

Social and cultural dynamics influence engagement. Group-based tasting events and storytelling improve authenticity and comfort. Tailoring experiences to gender and age differences increases satisfaction and broadens market reach.  

Future research should examine the long-term impact of culturally adaptive tasting experiences on consumer loyalty and winery performance. Cross-cultural tasting frameworks and sensory tools tailored to different markets could further enhance global engagement with wine tourism.  

\end{document}